\documentclass[reprint,eqsecnum,floats,aps,amsmath,amssymb,nofootinbib,prd,onecolumn,showpacs]{revtex4-2}

\usepackage{graphicx}
\usepackage{bm}
\usepackage{amsmath}	
\usepackage{amsfonts}
\usepackage{mathtools}
\usepackage{amssymb}							
\usepackage{braket}
\usepackage{amsthm}			
\usepackage{hyperref}

\usepackage[T1]{fontenc} 

\usepackage{color}

\begin{document}

\title{Analytic and Numerical Study of Scalar Perturbations in Loop Quantum Cosmology}

\author{Guillermo A. Mena Marug\'an}
\email{mena@iem.cfmac.csic.es}
\affiliation{Instituto de Estructura de la Materia, IEM-CSIC, Serrano 121, 28006 Madrid, Spain}

\author{Antonio Vicente-Becerril}
\email{antonio.vicente@iem.cfmac.csic.es}
\affiliation{Instituto de Estructura de la Materia, IEM-CSIC, Serrano 121, 28006 Madrid, Spain}

\author{Jes\'us Y\'ebana Carrilero}
\email{jesus.yebana@uib.es}
\affiliation{Departament de Física, Universitat de les Illes Balears, IAC3 – IEEC, Crta. Valldemossa km 7.5, E-07122 Palma, Spain}

\begin{abstract}
The possibility that quantum geometry effects may alleviate the apparent tensions existing at large angular scales in the observations of the Cosmic Microwave Background explains the increasing interest in considering primordial perturbations within the framework of Loop Quantum Cosmology. In this framework, a number of approximations have been suggested to simplify the study of perturbations and derive analytic expressions for the power spectra. This study requires a new choice of vacuum state that takes into account the preinflationary background geometry, choice for which we adopt the so-called NO-AHD prescription. Here, we apply the aforementioned approximations to the investigation of scalar perturbations. We discuss two approaches to the quantization of perturbations in Loop Quantum Cosmology, namely the hybrid and the dressed metric approaches. We improve previous approximations by including slow-roll corrections in the inflationary era. Moreover, for the first time in the literature of the NO-AHD prescription, we compute numerically the primordial power spectra for the two considered approaches, and show that the analytic estimations are remarkably accurate for all observable modes. We also discuss the similarities and differences between the spectra obtained with those two approaches.  
\end{abstract}

\maketitle

\section{Introduccion}

The Cosmological Microwave Background (CMB) is an important tool to explore the physics of the Early Universe \cite{CMB}. Among the observables available in the CMB, the angular power spectrum is specially important given the precision that we have attained in its observations and the amount of information that can be extracted from it \cite{WMAP,Planck,Planck-inf}. Although these observations have provided strong support to the standard inflationary models of General Relativity (GR), there exist anomalies at large angular scales that reveal some tensions with those models \cite{Planck, Planck_anomalies}. This is the case of the power suppression anomaly, the lensing amplitude anomaly, or the parity anomaly \cite{ASr,ASr2,AgSr,AgSr2}. It has been suggested that these tensions may indicate new physics arising from a non-standard inflationary evolution or from preinflationary epochs \cite{Pre_inflation, Inflation_multi}. This includes the possibility of quantum gravity effects in the very early stages of our Universe. In this context, quantum geometry corrections originated from Loop Quantum Gravity (LQG) have attracted an increasing interest. LQG is a nonperturbative program for the quantization of GR \cite{LQG,Thie}. Its application to cosmological scenarios, like those of relevance for the study of the CMB, is usually called Loop Quantum Cosmology (LQC) \cite{LQC}. One of the results of LQC is the existence of quantum states peaked on effective trajectories that avoid the cosmological singularity, replacing it with a bounce that connects a contracting cosmology with an expanding one \cite{LQC, APS}. These effective trajectories differ from cosmological solutions in GR only near the bounce (typically for matter energy densities approaching a critical density of Planck order, $\rho_c$, which is the maximum allowed in LQC and which is reached precisely at the bounce \cite{APS2}). 
 
Considerable attention has been paid to the possibility of extracting predictions in LQC for the angular power spectrum of temperature anisotropies in the CMB, with an eye on falsifying the theory with observations (see e.g. Refs. \cite{ASr,ASr2,AgSr,AgSr2,AAN,AM,hybridCMB,barrCMB,wangCMB}). For this goal, understanding the behavior and evolution of primordial scalar perturbations in LQC is crucial. These scalar perturbations contribute also to other observables, such as the E-mode polarization and the cross-correlation between E-modes and temperature. Furthermore, lensed scalar E-modes induce B-polarization, additional to the one produced by tensor perturbations. Therefore, missions designed to measure these other observables can provide valuable complementary information \cite{SPT,Bicep3,LiteBIRD}.

Primordial scalar perturbations can be described in terms of the so-called Mukhanov-Sasaki (MS) gauge invariant \cite{Mukhanov,Sasaki,Sasakikodama}, which follows a field equation that can be interpreted as a generalized wave equation with a background-dependent mass. This equation differs from the dynamical equation of tensor perturbations (written in a suitable form) by certain contributions of the inflaton potential to the mass \cite{Mukhanov1, Langlois, Baumann}. On the other hand, when LQC effects are considered, the mass is modified with quantum geometry corrections. The corrections do not coincide for tensor and scalar perturbations, i.e., the difference between the masses of these perturbations is changed by quantum contributions. As a consequence, the imprints transmitted to the primordial power spectrum (PPS) are not exactly the same for these two types of perturbations.   

Several approaches have been suggested within LQC to study primordial perturbations \cite{dressed1,dressed2,hybr_inf1,hybrid_rev,effective2,effective3,effective4,effective5}. 
Two of these approaches respect the hyperbolicity of the equations of the perturbations in the ultraviolet sector and do not distort the relativistic dispersion relations. The main effect of the quantum geometry is to modify the background-dependent mass \cite{NBMmass}. These are the hybrid \cite{hybridCMB,hybr_inf1,hybrid_rev,hybr_inf2,hybr_ref,hybr_ten} and the dressed metric \cite{AAN,AM,dressed1,dressed2,dressed4} approaches. The hybrid approach treats the whole system, composed of the background and the perturbations, as a constrained canonical system \cite{hybrid_rev}. The resulting Hamiltonian constraint is quantized combining a loop quantization of the background with a conventional Fock quantization of the perturbations. Using a kind of mean field approximation, one can extract propagation equations for the perturbations from this constraint. For quantum states of the background that are peaked on an effective trajectory of LQC \cite{APS2,Taveras}, these equations for the perturbations reproduce the equations of GR except for changes in the mass term, which is now obtained by evaluating the canonical background variables on the effective solution described by the peak \cite{hybrid_rev,NBMmass}. In contrast, in the dressed metric approach one first quantizes the background geometry, considers peaked states, assigns to their peaks a metric that is dressed with quantum corrections, and lifts this metric to the phase space of the perturbations \cite{AAN,dressed2}. The resulting propagation equations differ again from those of GR inasmuch as the dressed metric is not a classical relativistic solution. Furthermore, they also differ from the hybrid equations because the classical relations between canonical momenta and time derivatives are not respected by the quantum modifications encapsulated in the effective dynamics of LQC, which governs the evolution of the dressed metric \cite{NBMmass}. Consequently, in the regime where quantum effects are important, the two approaches result in distinct background-dependent masses for the perturbations.

In order to determine the cosmological perturbations at the end of inflation, in addition to their propagation equations, we need initial conditions for them. These conditions are usually identified as corresponding to a natural vacuum state for the perturbations \cite{NBM}. In standard slow-roll inflation, a natural choice of vacuum is the Bunch-Davies state, invariant under the symmetries of de Sitter spacetime and with a good ultraviolet (Hadamard) behavior \cite{Bunch}. In LQC scenarios, however, we must not expect that the Bunch-Davies state be a reasonable choice of vacuum state, at least for modes able to feel the quantum effects of the preinflationary stages near the bounce. These modes have wavelengths in those epochs of the order of the Planck scale \cite{AAN}. A vacuum that is specially well-adapted to the background dynamics in these situations is the NO-AHD (from the initials of Non-Oscillating with Asymptotic Hamiltonian Diagonalization) state \cite{NMT,NMP}. The modes of this state diagonalize the Hamiltonian of the perturbations in the asymptotic limit of infinitely large wavenumbers (ultraviolet modes). In addition, the state leads to a power spectrum without rapid oscillations, suppressing spurious power contributions that could arise from an average of those oscillating contributions \cite{NM}. In this work, we will adopt this choice of vacuum state.   

The numerical evolution of the perturbations and the determination of the power spectrum when that evolution freezes is demanding and time consuming. Owing to this, it is very convenient to develop approximation methods to estimate the results in an analytic manner. Such approximations have been proposed in the literature, first for the hybrid approach \cite{NM} and more recently for the dressed metric case \cite{AMV}. The strategy consists of dividing the evolution from the bounce until the end of inflation in several intervals where one can approximate the background-dependent mass in such a way that an analytic resolution of the equations is possible. The approximations are tailored to the behavior of the effective LQC backgrounds that have phenomenological interest in the discussion of the CMB. These backgrounds suffer a number of e-folds compatible with observations and lead to quantum corrections in the power spectrum that can affect the modes measurable today \cite{AM}. It turns out that these backgrounds have an inflaton energy density at the bounce dominated by the kinetic contribution, and experiment short-lived inflation \cite{AM,NBM,NM,waco}.

Strictly speaking, the aforementioned approximations were originally conceived for tensor perturbations, but are applicable to scalar perturbations at the price of ignoring the influence of the potential, except in order to sustain a de Sitter inflation. In the case of the hybrid approach, however, the approximations have recently been improved in Ref. \cite{NMY} to include the effect of the inflaton potential, both at first order during kinetic domination (in a kind of perturbative treatment of this potential) and in slow roll during inflation. In the present work, and still adhering to the NO-AHD proposal to select the vacuum state, we are going to consider exclusively scalar perturbations and adapt the available approximations to simplify the analytic description as much as possible, although retaining slow-roll corrections in the inflationary regime to discuss the dependence on the inflaton potential. We will apply these improved approximations not only to the hybrid approach, but also to the dressed metric approach, completing the study of the scalar perturbations and of the analytic evaluation of their power spectrum in both cases. Furthermore, for the first time in the literature, we will compute numerically the PPS for both approaches adopting the NO-AHD vacuum. This will allow us to check the goodness of our approximations and demonstrate that they capture the most relevant aspects of the dynamics of the perturbations. Indeed, we will prove that the analytic estimations are remarkably accurate. We will also compare the power spectra obtained with each of the two considered quantization approaches within LQC. For convenience, in this work, we will carry out our analytic and numerical calculations using a quadratic inflaton potential, but our study can be generalized to other potentials.

The structure of the rest of this paper is as follows. In Sec. 2 we present the class of backgrounds of interest in effective LQC that we are going to consider, and the initial conditions on the perturbations corresponding to the NO-AHD state. Then, we study the corresponding background-dependent mass, its approximations, and the exact resolution of the mode equations for the scalar perturbations, both for the hybrid and the dressed metric approaches. Sec. 3 is devoted to the analytic and numerical computation of the PPS. This numerical computation is novel in LQC for the NO-AHD vacuum. We compare the resulting spectra, studying also the two considered approaches. Finally, Sec. 4 contains the conclusions. Details of the numerics are described in the Appendix. We use Planck units in our calculations, setting $G$, $c$, and $\hbar$ equal to one.

\section{Background-dependent mass}

In this section, we summarize the most relevant results about the dynamics of the effective backgrounds in LQC with phenomenological interest for the CMB and about the background-dependent mass of the scalar perturbations corresponding to such backgrounds in the hybrid and the dressed metric approaches (for further details, see e.g. Refs. \cite{AM,hybrid_rev,NBMmass}). We also apply and improve the approximations to this mass discussed in Refs. \cite{NM,AMV}.

\subsection{The background}

We consider backgrounds that possess a homogeneous and isotropic Friedmann-Lemaître-Robertson-Walker (FLRW) metric and contain a homogeneous scalar inflaton field $\phi(t)$, subjected to a potential $V(\phi)$, which we particularize for convenience to a quadratic potential of the form $m^2\phi^2/2$. This inflaton potential can be generalized to other cases without serious obstructions, and even our analytic calculations can be reproduced for certain families of polynomic or exponential potentials, including the Starobinsky potential \cite{Simon}. Motivated by phenomenological reasons, in order to get results that include reasonable quantum corrections while retaining compatibility with observational data \cite{AM,NM}, and also for easy comparison with previous computations available in the literature \cite{AAN,NBM,NM,NMY}, we take an inflaton mass equal to $m=1.2 \times 10^{-6}$ in Planck units. The energy density and pressure of the scalar field are $\rho = V(\phi)+ (\phi^{\prime})^2/(2a^2)$ and $P =\rho - 2 V(\phi)$, respectively, where the prime denotes the derivative with respect to conformal time. We reserve the dot symbol for the proper time derivative. The background dynamics in effective LQC is given by the equations \cite{AAN,hybridCMB}
\begin{eqnarray}\label{eq_Friedman_LQC}
    \left( \frac{a'}{a}\right)^{2} = \frac{8\pi}{3}a^{2}\rho \left(1 - \frac{\rho}{\rho_{c}} \right),\quad \frac{a''}{a} = \frac{4\pi}{3}a^{2}\rho \left(1 + 2\frac{\rho}{\rho_{c}} \right) - 4\pi a^{2}P \left(1 - 2\frac{\rho}{\rho_{c}} \right).
\end{eqnarray}
The first one is a modified Friedmann equation that can be treated as a constraint in (unperturbed) effective LQC. The second equation can be obtained from this first one by derivation, using the local conservation law for the inflaton energy density.

As we mentioned in the Introduction, the inflaton energy density reaches a maximum at the bounce, equal to $\rho_c = 3 /(8\pi \gamma^2 \Delta)$, where $\gamma$ is the Immirzi parameter \cite{immirzi} and $\Delta = 4 \sqrt{3}\pi \gamma $ is the area gap, derived from the area spectrum of LQG \cite{LQG,Thie}. Here, we take the usual value $\gamma=0.2375$, motivated by black-hole entropy calculations in LQG \cite{LQG}. We can integrate the background dynamics starting at the bounce, where the derivative of the scale factor vanishes, and set the scale factor equal to one there, as a reference scale. The modified Friedmann equation provides then a relation between the initial value of the inflaton and its conformal time derivative. Hence, we only need to choose an initial value for our inflaton field. For concreteness, we take $\phi_0 = 0.97$, a value that leads to effective background solutions of phenomenological interest and has been studied in several previous analyses \cite{AM,NBM,NM}, in consonance with our comments above about the inflaton mass. The results are not qualitatively affected if we slightly change this initial value.

\subsection{Approximate analytic solution for the hybrid approach}
 
In the case of the hybrid approach to the quantization of scalar perturbations within LQC, the background-dependent mass $ s^{(s)}$ has the form \cite{NBMmass,hybrid_rev}
\begin{eqnarray} \label{eq_hyb_mass}
    s^{(s)} =  - \frac{4\pi }{3} a^2 (\rho-3P) + \mathcal{U}, \qquad  \quad      \mathcal{U}= a^2  \left[ V_{,\phi \phi} + 48\pi V + 6\frac{a' \phi '}{a^{3}\rho}V_{,\phi} - \frac{48\pi}{\rho}V^{2} \right].
\end{eqnarray}
The comma in the potential stands for the derivatives with respect to the inflaton. The Fourier modes of the scalar perturbations satisfy the equation $\mu_{k}^{\prime\prime} + (k^2 + s^{(s)} ) \mu_{k} = 0$, where $k$ is the (angular) wavenumber of the mode.

Two facts are crucial to introduce convenient approximations to the above mass. First, the mass is almost indistinguishable from its counterpart in GR away from the region surrounding the bounce. In other words, quantum geometry effects are only relevant during a short interval near the bounce. Second, the influence of the inflaton potential is important only after the onset of inflation, once the epoch of kinetic domination has finished. 

This kinetic era can be subdivided into two, namely, the aforementioned interval with non-negligible quantum corrections around the bounce, and a subsequent classical relativistic period with kinetically dominated evolution. During these two periods, we can ignore the presence of the inflaton potential, for backgrounds of interest in effective LQC. Therefore, the mass of the scalar perturbations can be treated in these periods with the same approximations discussed in Refs. \cite{NM,AMV} as if they were tensor perturbations. Kinetic domination is followed by an inflationary era, where the potential plays a key role. Even though very close to the onset of inflation the expansion is not yet in a slow-roll regime (in a restricted period when the evolution transitions from kinetic domination to inflation), the slow-roll approximation is quite successful in incorporating the global effect of inflation on the perturbations. In particular, it reflects the main imprint of the inflaton potential on the perturbation modes before their dynamics freeze. Here, we include such slow-roll corrections to the description of the scalar perturbations, in contrast with the original approximation made in Ref. \cite{NM} where these corrections were ignored. These corrections are essential to understand the influence of the inflaton potential in the PPS and ultimately discuss whether a specific potential is favored by observational data.

Let start with the bounce epoch. In this first interval, a Pöschl-Teller potential provides a good approximation to the effective background-dependent mass. The parameters of the Pöschl-Teller potential are fixed with conditions at the bounce and at the end of the interval, as it is done in Refs. \cite{NM,AMV}. Also, as in those references, the end of the interval is chosen to optimize the approximation to the mass in the union of this bounce era and the following classical interval with kinetic domination. For our initial condition on the inflaton, this leads to a proper time of $t_0=0.4$ Planck units \cite{NM}. The exact general solution to the mode equation for the scalar perturbations with a Pöschl-Teller mass is known analytically. We have to select the specific solution that corresponds to our choice of vacuum state. As commented in the Introduction, a Bunch-Davies state ceases to be a natural choice of vacuum when we consider the preinflationary dynamics of LQC and, in particular, the bounce. Other possible choices of vacuum have been proposed in LQC \cite{Agullo1,gupt,gupt2,deBlas}. In this work, we adhere to the NO-AHD proposal \cite{NMT}, which selects a state with modes that diagonalize the Hamiltonian of the perturbations in the ultraviolet (and ultimately must lead to a power spectrum without rapid oscillations). With the Pöschl-Teller approximation and imposing the asymptotic diagonalization condition, restricted to the bounce epoch, it is possible to determine the vacuum state, as shown in Ref. \cite{NM}. This state determines in turn the following (normalized) mode solutions:
\begin{equation} \label{PTEq}
\mu_{k}^{PT} =\frac{1}{\sqrt{2k}}\left[ x \left( 1-x\right) \right]^{-\frac{ik}{2\alpha}} {}_{2}F_{1}\left( c_{+}-\frac{ik}{\alpha}, c_{-}-\frac{ik}{\alpha}, 1-\frac{ik}{\alpha}; x \right),
\end{equation}
where we have chosen the origin of time at the bounce, ${}_{2}F_{1}$ is a hypergeometric function, and
\begin{equation}
x = \frac{1}{1 + e^{-2\alpha \eta} },  \qquad \alpha =\frac{\mathrm{arcosh}{(a_{0}^{2})}}{\eta_{0}^{2}}, \qquad c_{\pm} = \frac{1}{2} \left( 1 \pm \sqrt{1 + \frac{32\pi\rho_{c}}{3\alpha^{2}}} \right).
\end{equation}
Here, $\eta_0$ is the conformal time at the end of the bounce interval (corresponding to the proper time $t_0$) and $a_0$ is the scale factor at that moment. Evaluating this solution and its derivative at $\eta_0$ and requiring continuity in the modes up to first derivatives, we then obtain initial values to integrate the modes in the classical period following the bounce. 

This second interval is characterized by kinetically dominated dynamics. Thus, the influence of the potential can be ignored. In addition, quantum effects are already negligible, and the study can be carried out as in GR. Therefore, the mass of the scalar perturbations can be approximated as in the tensor case with kinetic domination \cite{NM}, namely, $ s^{(s)}= s_\text{GR}^{(s)}$ with
\begin{eqnarray}
s_\text{GR}^{(s)} (\eta) = \frac{1}{4 {\bar{y}}^2}, \qquad \bar{y}=\eta-\eta_0 + \left(\frac{1}{2H_0a_0 }\right) , \label{eq_Effective_mass_GR}
\end{eqnarray} 
where $H_0$ is the Hubble parameter at the initial time of this interval, $\eta_0$. With the above mass, the general solution to the mode equation is 
\begin{eqnarray}
\mu_{k}^{GR} (\bar{y}) = \sqrt{\frac{\pi \bar{y}}{4}} \Big[C_k H_0^{(1)} (k\bar{y}) + D_k H_0^{(2)}(k\bar{y}) \Big ],\label{eq_sol_pert_kin}
\end{eqnarray}
where $H_{0}^{(1)}$ and $H_{0}^{(2)}$ are the Hankel functions of zeroth order and of the first and second kind, respectively. The integration constants $C_k$ and $D_k$ are fixed by our continuity requirements with the mode solution $\mu_k^{PT}$ at the bounce (Pöschl-Teller) period. Calling $\hat{k}=k/(2a_0H_0)$, we obtain \cite{NM}
\begin{eqnarray}\label{CD}
C_{k} &=&  \frac{1}{H_{0}^{(1)}(\hat{k})} \left[\sqrt{\frac{8 H_0a_0}{\pi}} \mu_{k}^{PT}(\eta_0) - D_{k} H_{0}^{(2)}(\hat{k})\right],\\
D_{k} &=& i\sqrt{\frac{\pi}{8 H_0a_0}}\left[ k H_{1}^{(1)}(\hat{k}) \mu^{PT}_{k}(\eta_0) - H_0 a_0 H_{0}^{(1)}(\hat{k})  \mu^{PT}_{k}(\eta_0)
+ H_{0}^{(1)}(\hat{k}) \dot{\mu}^{PT}_{k}(\eta_0) \right].
\end{eqnarray}

\begin{figure}
    \centering
    \includegraphics[width=16cm]{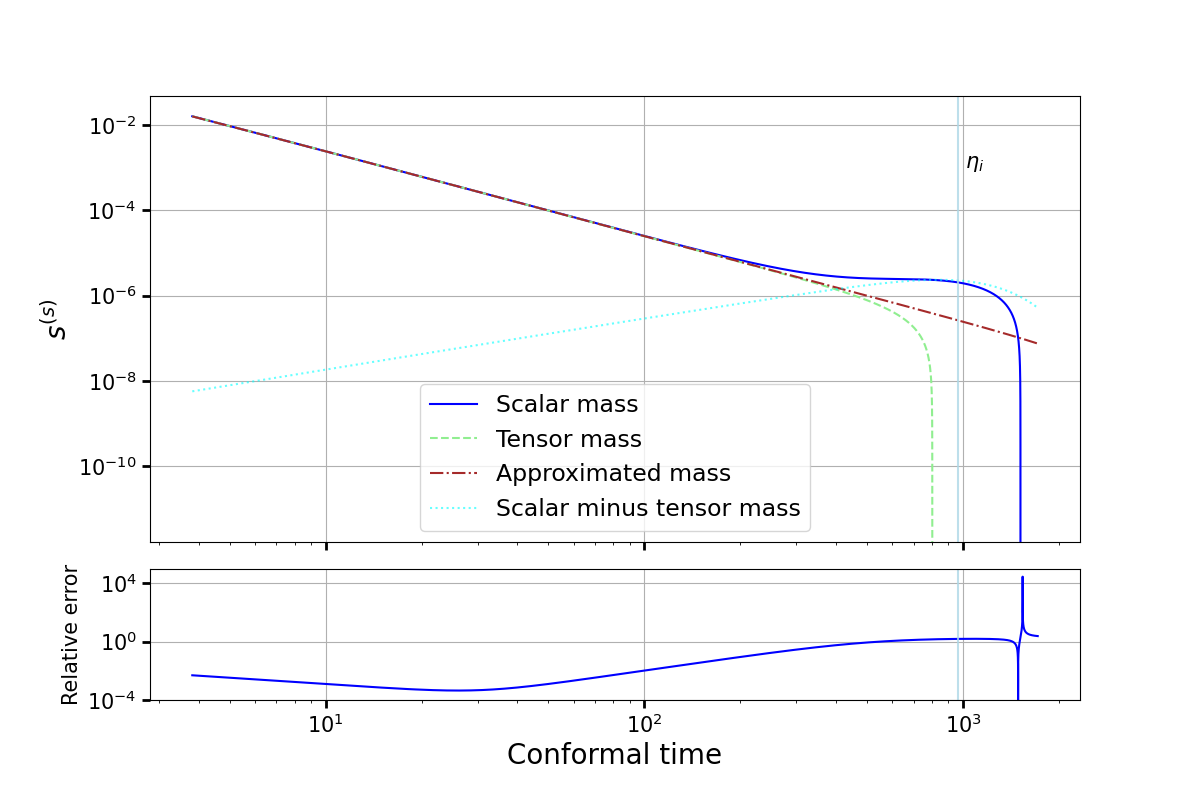}
    \caption{Top: Numerical computation of the background-dependent mass $s^{(s)}$ for scalar perturbations (continuous blue line) and tensor perturbations (dashed green line) in the hybrid approach, its numerical difference (dotted cyan line), and the approximated mass obtained using a kinetically dominated regime in GR (dash-dotted red line). This background-dependent mass is defined in Eq. \eqref{eq_hyb_mass}, with vanishing potential contribution $\mathcal{U}$ for tensor perturbations, and its approximation is given by Eq. \eqref{eq_Effective_mass_GR}. Bottom: Relative error between the numerical and the approximated values of the mass for scalar perturbations. We see that the approximation to the background-dependent mass is very good, with a relative error below 0.1 in most of the considered interval and which is always below 1.6, including conformal times close to the instant $\eta_i$ corresponding to the beginning of inflation, where this mass is actually very small. We have taken $\gamma=0.2375$ and $\phi_0=0.97$ for the Immirzi parameter and the value of the inflaton at the bounce, respectively, and considered a quadratic inflaton potential with mass $m=1.2 \times 10^{-6}$.} 
    \label{fig_Masa_Hyb_S_GR}
\end{figure}

We place the beginning of the inflationary era when the inflaton potential contribution to the energy density starts to be of the same order as the kinetic term. We choose the transition time to ensure a good approximation to the background-dependent mass. More specifically, we set this time at $\eta_i= 966$, before the instant when the relative error between the exact mass and the approximated mass in kinetic domination starts to grow rapidly (see Fig. \ref{fig_Masa_Hyb_S_GR}). It is worth commenting that this choice, based on our numerical analysis of the mass, differs appreciably from the corresponding one in the fit of the mass for tensor perturbations (discussed in detail in Ref. \cite{Paper}). This difference is due to the introduction of the new potential term $\mathcal{U}$ in the scalar mass, which maintains its positivity for a longer interval than in the tensor case.

Finally, we consider the inflationary period in slow-roll approximation. The relevant slow-roll parameters are \cite{Langlois, Baumann} $\varepsilon_V = V_{,\phi}^2/(16\pi V^2)$ and $\delta_V = V_{,\phi\phi}/(8\pi V)$. Defining $\nu= \sqrt{9+ 36\varepsilon_V- 12\delta_V}/2 $, the slow-roll approximation to the mass of the perturbations yields \cite{Baumann, NMY}\begin{eqnarray}
        s^{(s)}_{SR}  =  -\frac{\nu^2 - \frac{1}{4}}{(\eta_{e}-\eta)^2},
\end{eqnarray}
where $\eta_e$ signals the end of inflation. The general mode solution with this mass is \cite{Baumann,NMY}
    \begin{equation} \label{eq_sol_MS_SR}
        \mu_k^{SR}(\eta) = \sqrt{\frac{\pi}{4}(\eta_{e}-\eta)} \left\lbrace A_k H_\nu^{(1)} \left[k(\eta_{e}-\eta)\right] + B_k H_\nu^{(2)} \left[k(\eta_{e}-\eta)\right]\right\rbrace,
    \end{equation}
where $H_{\nu}^{(1)}$ and $H_{\nu}^{(2)}$ are Hankel functions, now of order $\nu$. The integration constants $A_{k}$ and $B_{k}$ can be calculated by imposing continuity of the modes up to the first derivative at the matching point $\eta_i$ with kinetic domination, getting
\begin{equation}
A_{k} =  -i\sqrt{\frac{\pi}{16}\Delta\eta} \Bigg[ kH_{\nu+1}^{(2)}\left(k\Delta\eta\right) 
- kH_{\nu-1}^{(2)}\left(k\Delta\eta\right) - \frac{H_{\nu}^{(2)}\left(k\Delta\eta\right)}{\Delta\eta} \Bigg] \mu_{k}^{GR}(\eta_{i})
+i\sqrt{\frac{\pi}{4}\Delta\eta}H_{\nu}^{(2)}\left(k\Delta\eta\right) \dot{\mu}_{k}^{GR}(\eta_{i}), \end{equation}
and a similar equation for $B_{k}$ except for a global minus sign and the interchange of first order Hankel functions by second order ones and vice versa. 
Here, $\Delta\eta=\eta_e-\eta_i$. On the other hand, normalization of the mode solutions requires that $\mu_k^{SR} (\mu_k^{SR})^{*\prime}-(\mu_k^{SR})^{\prime}(\mu_k^{SR})^{*}=1$ which, in terms of the integration constants, amounts to the condition $|B_k|^2=1+|A_k|^2$. The symbol $*$ stands for complex conjugation. It is possible to check that this condition is automatically satisfied if the mode solutions have been normalized in any of the intervals preceding the inflationary period, as we did. Then, substituting the above expressions in the slow-roll solution, we obtain the scalar modes. Their evaluation when the evolution freezes provides the PPS. 

\subsection{Approximate analytic solution for the dressed metric approach}

The dressed metric approach incorporates the most important quantum effects on the homogeneous background in a metric corrected with quantum contributions, and then describes the evolution of the perturbations as test fields propagating on this dressed background. The background-dependent mass for scalar perturbations on effective solutions to LQC has the form \cite{NBM} 
 \begin{eqnarray} \label{eq_dress_mass}
     s^{(s)} = - \frac{4\pi}{3}a^{2}\rho \left(1 + 2\frac{\rho}{\rho_{c}} \right) + 4\pi a^{2}P \left(1 - 2\frac{\rho}{\rho_{c}}   \right) + \mathcal{V}, \end{eqnarray}
where
\begin{eqnarray}
    \mathcal{V} = a^2 \left[V_{,\phi\phi} +48\pi  V - \frac{48\pi }{\rho} V^2 - \sigma_a \sqrt{\frac{96\pi }{\rho}} \frac{|\phi'|V_{,\phi}}{a} \right]. 
\end{eqnarray}
Here $\sigma_a$ is the sign of the product between $\phi'$ and the canonical momentum of the scale factor. This expression differs from its counterpart in the hybrid case. The difference is significant in the interval around the bounce, where quantum effects are relevant, and dilutes becoming negligible as soon as the dynamics enter a classical regime in the interval of kinetic domination. Therefore, the mass of the scalar perturbations can be treated in the same way and with the same approximations as in the hybrid analysis after the bounce period. Obviously, this happens also for the mode solutions. In the bounce interval, on the other hand, the type of Pöschl-Teller approximation employed in the hybrid approach is still applicable if we change the parameters of the Pöschl-Teller potential and add a constant to it. This constant adapts the approximation to the fact that the mass of the dressed metric case turns out to change sign during the stages with important quantum effects. It starts being negative at the bounce and flips signs before the classical regime with kinetic domination starts. Our procedure allows us to extend the NO-AHD proposal to the dressed metric approach. This is of utmost importance if we want to compare the influence of the quantization approach used in LQC while adhering to the same criterion for the choice of vacuum. It also serves to avoid prejudicies favoring other vacuum proposals \cite{ASr,ASr2,AgSr,AgSr2} as more suited to the dressed metric approach. The inclusion of the additive constant in the Pöschl-Teller mass does not change much the analytic solution for the scalar modes: it only implies a redefinition of the wavenumber and the variable $x$ that appear in Eq. \eqref{PTEq}, and a change of the constants $c_{\pm}$ in that equation. Actually, since the influence of the inflaton potential is negligible in the bounce period, the resulting formula for the mode solution can be found in Ref. \cite{AMV}, where the case of tensor perturbations in the dressed metric approach was discussed. 
 
The main difference in the inflationary stages with respect to the hybrid approach is found in the specific values of the constants $A_k$ and $B_k$ for the scalar mode solutions. These constants are fixed by the continuous matching (up to first derivatives) of the solutions through the different evolution intervals, backward until we reach the bounce period where the initial conditions for the modes are set by the choice of an NO-AHD vacuum state. Therefore, changes in those initial conditions involve a subsequent change of $A_k$ and $B_k$, which are the constants that determine the PPS when the evolution of the modes freezes during inflation.

\section{Primordial Power spectrum}

With our approximated analytic mode solutions, we can now estimate the PPS. For this, we only need to evaluate the amplitude of the solutions at the moment $\eta_{f}$ when the variation of all relevant modes has frozen. This happens approximately when the expansion has reached 30 e-folds \cite{NBM}. The PPS of the scalar perturbations is given by \cite{Langlois, Baumann}
\begin{eqnarray}\label{eq_def_PPS}
        \mathcal{P}_\text{R} = \frac{k^{3}}{2\pi^2} \frac{|u_k(\eta_{f})|^2}{z(\eta_{f})^2},
\end{eqnarray}
where $z=a^2\phi'/a'$, which in the slow-roll regime becomes $z^2 = a^2 \varepsilon_V/(4\pi)$. We can also use the slow-roll approximation to rewrite the mode solution \eqref{eq_sol_MS_SR} in a much easier form, because the argument of the Hankel functions is much smaller than one, at least for wavenumbers in a sufficiently large range which includes the observable window. We employ the asymptotic behavior of the Hankel functions \cite{Abra} to rewrite
\begin{eqnarray}
        \left|\mu_k^{SR}\right| = \frac{1}{4\pi} (\eta_{e}-\eta_{f}) |\Gamma(\nu)|^2 \left[\frac{k(\eta_{e}-\eta_{f})}{2}\right]^{-2\nu} |A_k - B_k|^2,
    \end{eqnarray}
where $\Gamma$ is the gamma function. The corresponding analytic expression of the PPS is $ \mathcal{P}_\text{R} = C_\nu k^{3-2\nu} |A_k-B_k|^2$, where
\begin{eqnarray} \label{eq_PPS_escalar_SR}
      C_\nu = \frac{1}{2\pi^2} \frac{\eta_{e}-\eta_{f}}{\varepsilon_V a_{f}^2 }  |\Gamma(\nu)|^2  \left(\frac{\eta_{e}-\eta_{f}}{2}\right)^{-2\nu}.
\end{eqnarray}
Here, $a_{f}$ is the scale factor at the conformal time $\eta_{f}$.

Even if we have determined our mode solutions starting with a state selected by the asymptotic Hamiltonian diagonalization condition, the fact that we needed approximations to characterize it analytically, including that we restricted first our considerations to the bounce period and adopted a Pöschl-Teller potential there, introduces spurious oscillations, accumulated during the whole interval of evolution. However, these spurious oscillations are now easy to remove, as explained in Ref. \cite{NM}, arriving at last to a genuine NO-AHD state. This can be accomplished by a Bogoliubov transformation which eliminates the rapidly varying phases of the integration constants $A_k$ and $B_k$. Note that the absolute values of these constants also vary with the wavenumber, but this change is smooth and slow, permitting a clear distinction of variation scales with respect to the fast oscillations of the phase. The desired  Bogoliubov transformation (that indeed respects the normalization condition $|B_k|^2=1+|A_k|^2$) can be directly implemented by means of the correspondence 
\begin{eqnarray}
A_k \to \Tilde{A}_k = |A_k|, \quad \quad B_k \to \Tilde{B}_k = |B_k|.
\end{eqnarray}
The analytic expression for the NO-AHD PPS is thus
\begin{eqnarray}
\mathcal{P}_\text{R}(k) = C_\nu k^{3-2\nu} \left(|A_k| - |B_k|\right)^2 .\label{eq_PW_formula_cstes}
\end{eqnarray}
This expression retains the main information about the scale dependence of the original spectrum that was not a superimposed rapid oscillation. It is easy to realize that this PPS is the envelope of the minima of the oscillating one. 

On the other hand, we need to test the goodness of this analytical approximation to the PPS to be sure of its reliability. In order to achieve this goal, we have numerically calculated the solutions to the mode equation of the scalar perturbations, and evaluated the PPS with these solutions. This is the first numerical investigation of these questions in LQC with the NO-AHD proposal. Our numerical computations start just after the bounce period, at the time when the kinetic period begins, therefore employing the Pöschl-Teller approximation around the bounce. The reason why our numerical integration relies on this first approximated period is because we are able to determine explicitly the state of the perturbations selected by the NO-AHD prescription only by these means. The value of the approximated mode solutions for this choice of vacuum at the end of the short interval with quantum effects is taken as our initial data for the numerics.  This numerical integration is performed over the whole of the dynamical evolution until the moment when the relevant modes freeze. In the numerical case, the fact that one does not need to complete the evolution until the end of inflation, but just until the beginning of this freezing regime, is very important, because this considerably reduces the computation time. The only remaining step is to use the definition \eqref{eq_def_PPS} of the PPS to evaluate the spectrum with our numerical results. For convenience, we absorb the overall constant $C_\nu$ (see Eq. \eqref{eq_PPS_escalar_SR}) in a normalized version of the PPS. It is worth emphasizing that this PPS has been calculated starting from an asymptotic diagonalization of the Hamiltonian of the perturbations restricted to the bounce period, instead of considering the whole evolution interval, and it is, therefore, the numerical counterpart of our oscillating analytic spectrum, previous to the removal of spurious oscillations by means of an adjustment of the initial state via a suitable Bogoliubov transformation. More details about this numerical computation are given in the Appendix.

\begin{figure}[h!]
\includegraphics[width=16cm]{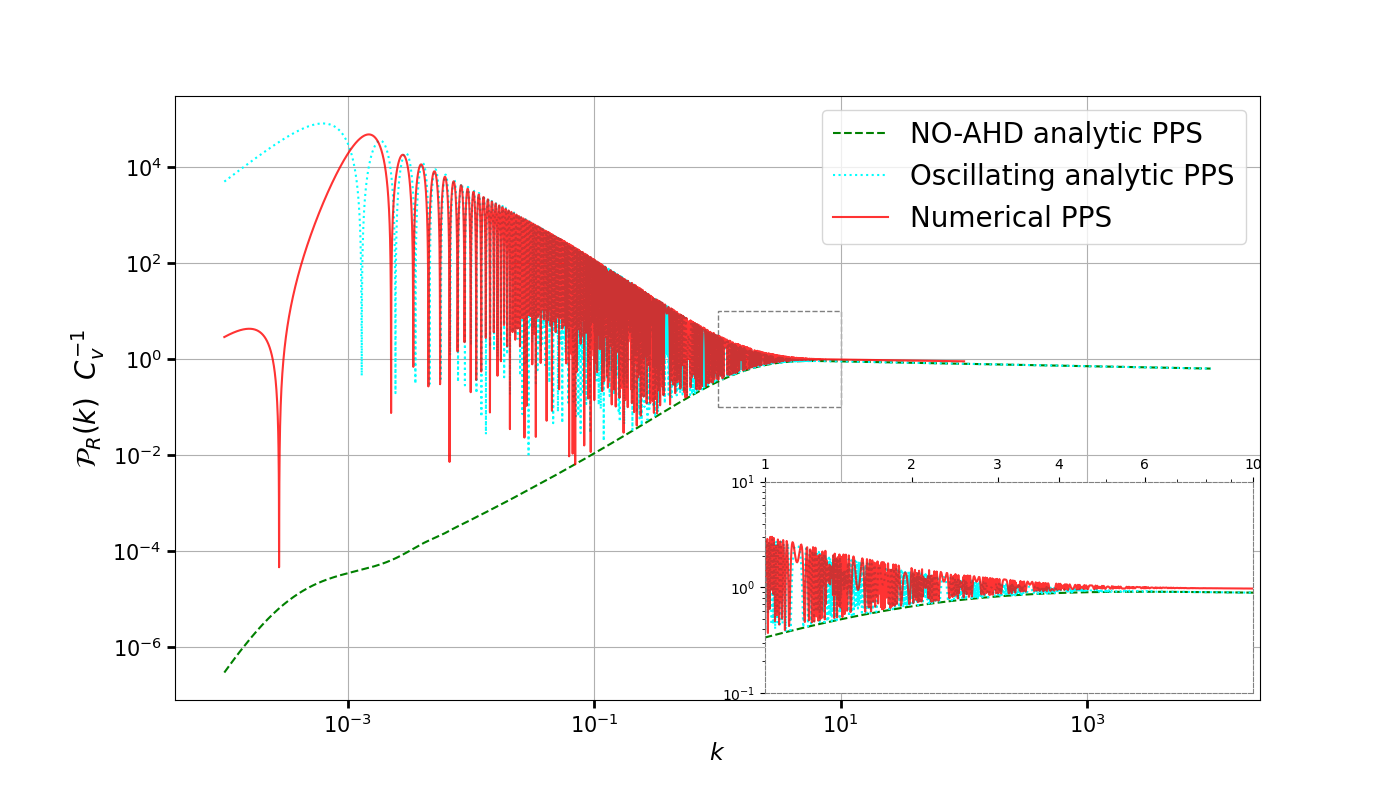}
\caption{Normalized primordial power spectrum (PPS) for the hybrid approach. We show the numerical computation of the oscillating PPS (continuous red line), the analytic approximation to this PPS  (dotted cyan line), and the analytic approximation to the PPS of the NO-AHD vacuum (dashed green line), obtained with a Bogoliubov transformation. We include an inset zooming in on the region $1\leq k \leq 10$ (framed in the PPS) to see the details when power suppression appears. We have taken $\gamma=0.2375$ and $\phi_0=0.97$ for the Immirzi parameter and the value of the inflaton at the bounce, respectively, and considered a quadratic inflaton potential with mass $m=1.2 \times 10^{-6}$.}
\label{fig_PPS_S_Hyb_Norma}
\end{figure}

\begin{figure}[h!]
\includegraphics[width=16cm]{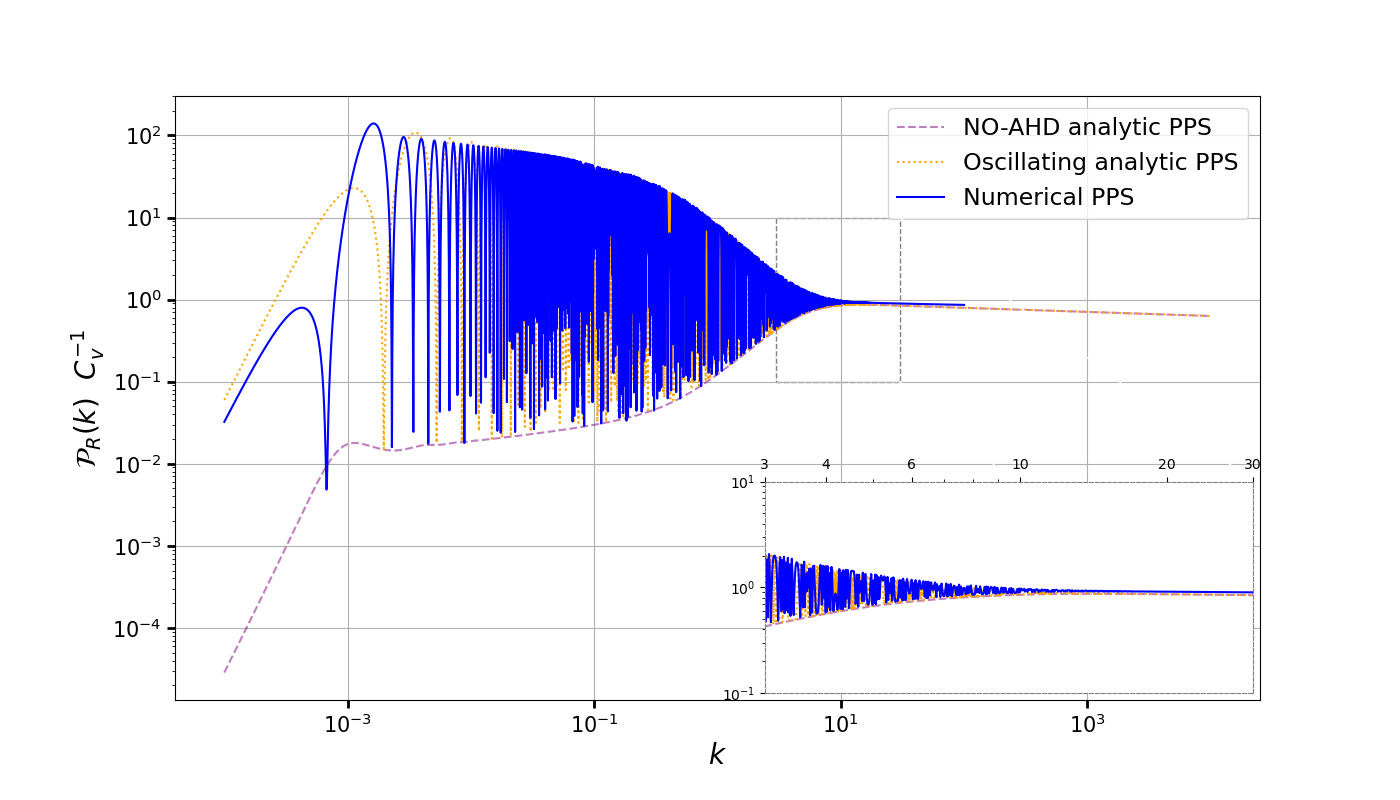}
\caption{Normalized primordial power spectrum (PPS) for the dressed metric approach. We show the numerical computation of the oscillating PPS (continuous blue line), the analytic approximation to this PPS (dotted orange line), and the analytic approximation to the PPS of the NO-AHD vacuum (dashed violet line), obtained with a Bogoliubov transformation. We include an inset zooming in on the region $3\leq k \leq 30$ (framed in the PPS) to see the details when power suppression appears. We have taken $\gamma=0.2375$ and $\phi_0=0.97$ for the Immirzi parameter and the value of the inflaton at the bounce, respectively, and considered a quadratic inflaton potential with mass $m=1.2 \times 10^{-6}$.}
\label{fig_PPS_S_DS_Norma}
\end{figure}

We can easily see that the NO-AHD analytic PPS for the hybrid case (obtained with the final Bogoliubov transformation) corresponds to the envelope of the minima of the oscillating PPS, as we commented above (see Fig. \ref{fig_PPS_S_Hyb_Norma}). In fact, we also see that this NO-AHD PPS fits very well the minima of the numerical PPS, both for the hybrid and the dressed metric approaches, displayed respectively in Figs. \ref{fig_PPS_S_Hyb_Norma} and \ref{fig_PPS_S_DS_Norma}. Hence, our numerical analysis demonstrates the remarkable goodness of our analytic approximations and prove that they capture the relevant physics. Nevertheless, the periods of the rapid oscillations seem slightly different in the infrared region when we compare the numerical and the oscillating analytic PPS. This discrepancy may come from the transition epoch between the kinetic dominated regime and the inflationary epoch, where our analytic approximation is less accurate. In addition, we see that far in the infrared and for the dressed metric approach, the match of the NO-AHD analytic PPS with the minima of the numerical PPS and of the oscillating analytic PPS is not evident. This might be a consequence of the fact that, in this region, the superimposed oscillations cease to vary much faster than the amplitudes $|A_k|$ and $|B_k|$ of the modes. In practice, however, this region is not relevant for observations, and the accuracy of the approximations is not quantitatively crucial there. In this sense, let us comment that the range of observable modes \cite{Planck} covers, in a rough estimation, the interval $10^{-4} \mathrm{Mpc}^{-1} \lesssim k/a_{today}\lesssim 10^{-1}\mathrm{Mpc}^{-1}$. This interval depends on the present value of the scale factor, $a_{today}$, which varies with the expansion experienced by the background, starting from a unit scale factor at the bounce. Those phenomenologically interesting background solutions with more e-folds can reach large values of the scale factor nowadays, with a corresponding lower bound for the window of observable modes ranging in the interval $[10^{-1}, 1]$ \cite{AM}, which excludes the far infrared region considered above. Let us also notice that our analytic and numerical PPS become indistinguishable at ultraviolet scales, where the tilt coming from the slow-roll inflationary process is clearly visible.      

\begin{figure} 
    \includegraphics[width=16cm]{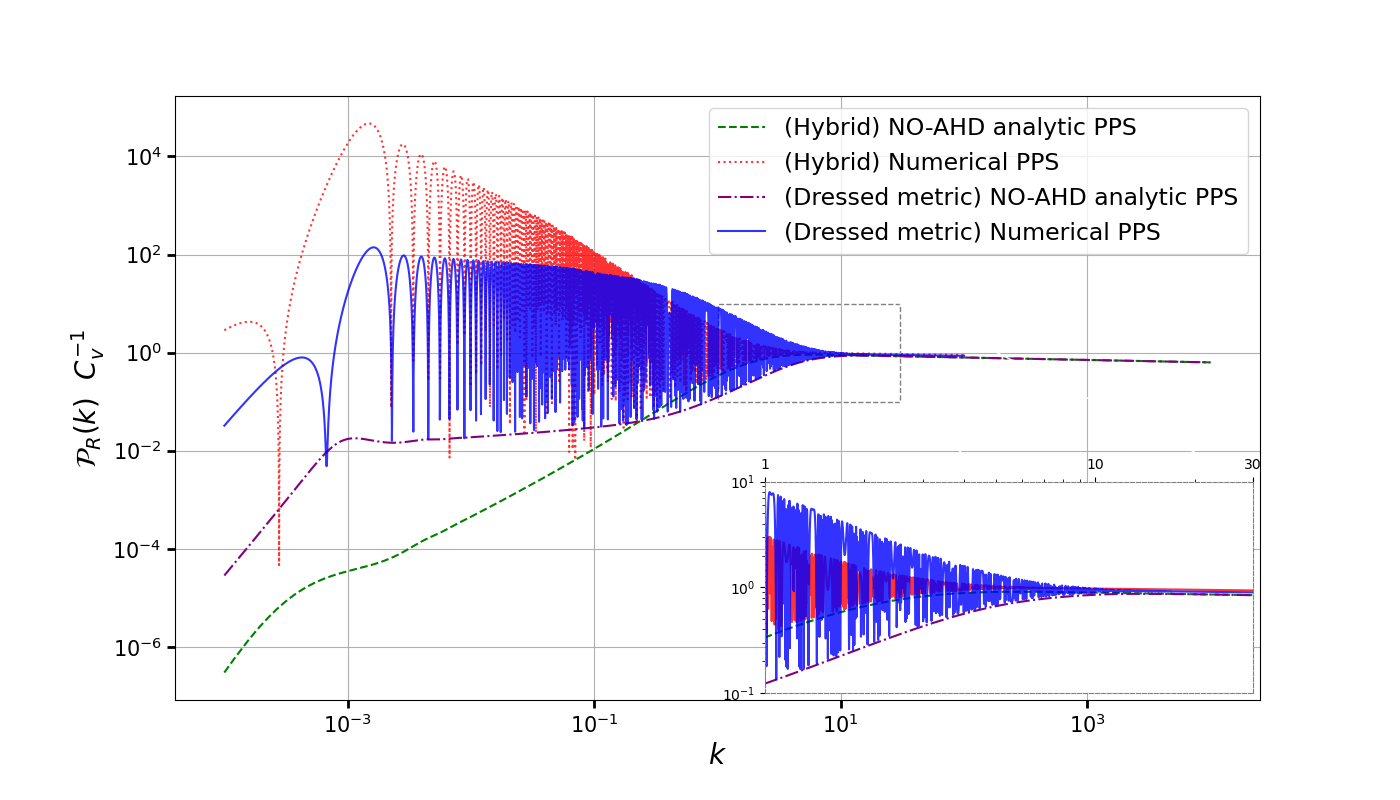}
    \caption{Comparison of the normalized primordial power spectra (PPS) for the hybrid and the dressed metric approaches. We show the numerical computation of the oscillating PPS for the hybrid approach (dotted red line) and for the dressed metric approach (continuous blue line), as well as the analytic approximations to the PPS of the NO-AHD vacuum for the hybrid approach (dashed green line) and the dressed metric approach (dash-dotted violet line). We include an inset zooming in on the region $1\leq k \leq 30$ (framed in the PPS) to see the details when power suppression appears. We have taken $\gamma=0.2375$ and $\phi_0=0.97$ for the Immirzi parameter and the value of the inflaton at the bounce, respectively, and considered a quadratic inflaton potential with mass $m=1.2 \times 10^{-6}$.}
    \label{fig_PPS_juntos}
    \end{figure} 
    
On the other hand, comparing the PPS of the two considered quantization approaches (see Fig. \ref{fig_PPS_juntos}), we see an effective cutoff with power suppression at wavenumbers that may well lie at the observable window, according to our previous discussion. This cutoff is not the same in both approaches. Actually, it is a bit larger in the dressed metric approach. However, we get more power suppression in the hybrid case, and this suppression is also steeper around the corresponding cutoff in this quantization approach. Finally, we emphasize that the PPS of the two approaches become almost identical for large wavenumbers, at low angular scales. 
        
\section{Discussion}

In this work, we have computed the PPS of scalar perturbations within the framework of LQC. We have investigated two different approaches to the quantization of primordial perturbations that respect the hyperbolicity of the field equations in the region of modes with large wavenumber. These are the hybrid and the dressed metric approaches. The field equations for the scalar perturbations reduce in both cases to (Fourier) mode equations of generalized harmonic oscillator type, with a background-dependent mass that differs in each of the two considered approaches. For backgrounds in effective LQC of phenomenological interest, in the sense that they are compatible with observations but may lead to observable modifications with respect to the standard inflationary predictions in GR, we have seen that the mass of the scalar perturbations remains positive in the case of the hybrid approach all the way from the bounce until the onset of inflation, while it starts in negative values at the bounce and then changes its sign in the dressed metric approach. 

The masses of the two approaches are indentical in practice as soon as the quantum geometry effects on the background can be ignored, something that happens soon after the bounce. Therefore, once this bounce period finishes and the dynamics enter the classical domain, the mass of the perturbations can be treated as coincident in hybrid and dressed metric LQC and in GR.  On the other hand, for the considered family of backgrounds in effective LQC with a suitable behavior, the influence of the inflaton potential is negligible until the beginning of the inflationary era. Consequently, both in the quantum bounce epoch, and in the classical regime until the onset of inflation, we can analyze the propagation of the perturbations ignoring the potential, adopting a kinetically dominated evolution. These considerations have allowed us to introduce a splitting of the evolution in different periods where we have been able to introduce manageable approximations for the background-dependent mass, leading in turn to analytic solutions to the mode equations in both of the two studied quantization approaches. Compared to previous analyses in the literature (see Refs. \cite{NM,AMV}), we have improved the transition times between the different periods (by benefitting from our new numerical calculations of the mass for the scalar perturbations) and introduced in both approaches a slow-roll approximation in the inflationary era which incorporates the effect of the inflaton potential, leading in particular to a tilt in the PPS for modes with large wavenumber.

The integration of the mode equations requires initial conditions for the perturbations, which are usually given in terms of a choice of vacuum state. Typically, in standard slow-roll inflation, this choice is provided by the Bunch-Davies state. However, the preinflationary dynamics of LQC and the existence of a bounce produce an evolution of the Hubble horizon radically different from that of slow-roll inflation, invalidating the privileged role of the Buch-Davies state in these scenarios with quantum effects, at least for modes with a wavenumber of the order of the curvature scale at the bounce. To resolve this problem, we have adhered to a recent proposal for the choice of a vacuum, based on an asymptotic diagonalization of the Hamiltonian of the perturbations at large wavenumbers. This proposal selects states without undesired superimposed oscillations. In fact, if this property is destroyed in the evolution by the impossibility of optimally adapting the selected state to all the details of the dynamics, it is possible to restore it and eliminate the spurious oscillations with a convenient Bogoliubov transformation, as we have shown. We have determined this so-called NO-AHD state with certain approximations around the bounce, and used it to provide initial conditions for the perturbations that have been evolved both analytically and numerically during the rest of the time span until the dynamics of the (relevant) modes freeze during the inflationary epoch. This numerical computation for the NO-AHD vacuum is totally novel in LQC.

As we have commented, in addition to our approximated methods, we have employed numerical integration to calculate the mode solutions and the corresponding PPS. The main conclusion of our study is that the analytic approximations provide a remarkably satisfactory estimation of the non-oscillating PPS for all relevant scalar modes, in both of the quantization approaches considered within LQC. Therefore, our numerical computation demonstrates that our analytic approximations can be used to obtain an excellent description of the PPS, avoiding demanding numerical simulations. Even so, before removing the undesired oscillations in the spectrum, there may exist some minor discrepancies in the oscillation frequency in the infrared region, which may be attributed to the transition from kinetic domination to inflation, which is the stage when our approximations are a little less accurate \cite{NMY}. On the other hand, our results also allow us to compare the spectra of the hybrid approach and the dressed metric approach. For both of them, we find power suppression at scales that may lie on the observable window. Notably, this power suppression is greater and steeper around the effective cutoff of the hybrid approach. Finally, for large wavenumbers, all the numerical and analytic spectra of the two approaches become almost indistinguishable, and reproduce the results of standard slow-roll inflationary models in GR, including a tilt that depends on the inflaton potential.

This work opens new avenues for future parametrizations of the PPS, dependent basically on the initial condition of the inflaton and the parameters of the background (like those in the inflaton potential or those determining the end of each regime in our approximations). This type of parametrization is essential for handling computations of the different power spectra and correlation functions of the perturbations, with an eye on their confrontation with observational data. In particular, this comparison should involve statistical analyses to find the best fittings and clarify the significance of the predictions. We consider the present work as a cornerstone for further research in this direction.

\acknowledgments

This work was supported by grants PID2020-118159GB-C41, PID2022-138626NB-I00, RED2022-134204-E, and RED2022-134411-T, funded by MCIN/AEI/10.13039/501100011033/FEDER, UE; the Universitat de les Illes Balears (UIB); the MCIN with funding from the European Union NextGenerationEU/PRTR (PRTR-C17.I1); the Comunitat Autonòma de les Illes Balears through the Direcció General de Recerca, Innovació I Transformació Digital with funds from the Tourist Stay Tax Law (PDR2020/11-ITS2017-006), the Conselleria d’Economia, Hisenda i Innovació grants No. SINCO2022/18146 and SINCO2022/6719, co-financed by EU and FEDER Operational Program 2021-2027 of the Balearic Islands; and the “ERDF A way of making Europe”. The authors are thankful to B. Elizaga Navascu\'es for important contributions in the development of this work. They are also thankful to A. Alonso-Serrano and P. Santo-Tomás for conversations. J.Y.C. is supported by the Spanish MICIU via an FPI doctoral grant (PRE2022-000809).

\appendix

\section{Numerical computation of the PPS} \label{Appendix}

The first step to numerically calculate the PPS is to integrate the evolution of the background and compute with it the exact value of the effective mass of the perturbations, both for the hybrid \ref{eq_hyb_mass} and the dressed metric \ref{eq_dress_mass} approaches. Since the scale factor changes very fast in conformal time during inflation, we integrate its dynamics in proper time, that requires less computational effort. Thus, we perform a numerical integration of the Friedmann equation and the energy conservation condition for the inflaton, from the bounce until the end of inflation. Actually, we split this numerical integration into two parts. Using a Runge-Kutta method of order 4, we first integrate with high precision around the bounce, in the interval $[0,100]$, and then we deal with the rest of the evolution employing an Adams-Bashforth-Moulton method, until the proper time $t_{e}= 10^{8}$ that approximately marks the end of the inflationary period. Inspecting the mass of the perturbations resulting from this integration of the background and comparing it with its analytic estimate allows us to improve in particular the transition times between the different stages of the dynamics, reducing the error of our approximations to the effective mass.

Once the background is determined numerically, we can proceed to integrate the mode equations for the dynamics of the perturbations. As we have commented above, we fix the initial state of the perturbations using a Pöschl-Teller approximation to the effective mass in a brief epoch around the bounce, shorter than a Planck second, imposing in this epoch the NO-AHD vacuum state, which can be found analytically. Then, we start the numerical integration with these data, beginning with the next epoch of kinetic domination. Following the same strategy used with the background, we integrate the mode equations in proper time. The conditions at the matching time $t_0$ between the bounce epoch and kinetic domination are
\begin{eqnarray}
\mu_{k}(t_{0}) = \sqrt{\frac{-1}{2\textrm{Im}h_{k}(t_{0})}}, \qquad \dot{\mu}_{k}(t_{0}) = \frac{-h_{k}^{*}(t_{0})\mu_{k}(t_{0})}{a(t_{0})},
\end{eqnarray}
where the function $h_{k}(t_0)$ corresponds to the NO-AHD vacuum at the considered time. Its detailed expression can be found in Ref. \cite{NM} and we do not repeat it here. Again, we divide the integration domain into two intervals, in order to improve numerical performance. The first one is chosen as $[t_0,3\times10^{5}]$, and approximately covers the kinetically dominated period, while the second interval is $[3\times10^{5},3.75\times10^{6}]$, and covers the inflationary regime until the moment when the evolution of all relevant modes is frozen. This happens around the time when the expansion reaches $30$ e-folds. For both intervals, we perform the integration taking $4\times10^7$ points. This precision is enough to have a good numerical result in each of the intervals. As a numerical check, the integration is carried out with both an Adams-Bashforth-Moulton method and a Runge-Kutta method of order 8. In both cases, we obtain very similar results. On the other hand, in principle we would have to perform the integration for each possible mode $k$. Note that the evolution of modes with large $k$ will be more numerically demanding, because they freeze later during inflation and therefore oscillate for longer periods of time than modes with small $k$. To get better control of this situation, we also split the integration into two different sets of $k$'s, namely those smaller or larger than unity, with the last set requiring a higher computational cost. Finally, the integration of all modes is performed with an absolute and relative tolerance of $10^{-9}$ in the integrators.

\end{document}